\documentclass[a4paper,aps,prd,10pt,preprintnumbers,showpacs,twocolumn,superscriptaddress,nofootinbib,amsmath,amssymb]{revtex4-1}
\usepackage[dvips]{graphics}
\usepackage[utf8]{inputenc}
\usepackage[T1]{fontenc}
\usepackage{cmap}

\begin{document}
\title{Analytical representation for metrics of scalarized Einstein-Maxwell black holes and their shadows}
\author{R. A. Konoplya}\email{konoplya\_roma@yahoo.com}
\affiliation{Institute of Physics and Research Centre of Theoretical Physics and Astrophysics, Faculty of Philosophy and Science, Silesian University in Opava, CZ-746 01 Opava, Czech Republic}
\affiliation{Peoples Friendship University of Russia (RUDN University), 6 Miklukho-Maklaya Street, Moscow 117198, Russian Federation}
\author{A. Zhidenko}\email{olexandr.zhydenko@ufabc.edu.br}
\affiliation{Institute of Physics and Research Centre of Theoretical Physics and Astrophysics, Faculty of Philosophy and Science, Silesian University in Opava, CZ-746 01 Opava, Czech Republic}
\affiliation{Centro de Matemática, Computação e Cognição (CMCC), Universidade Federal do ABC (UFABC),\\ Rua Abolição, CEP: 09210-180, Santo André, SP, Brazil}
\begin{abstract}
Here we construct approximate analytical forms for the metric coefficients and fields representing the scalarized Einstein-Maxwell black holes with various couplings of the scalar field, once the parameters of the system are fixed. By increasing approximation order, one can obtain the analytic representation with any desired accuracy, what was tested via calculations of shadows for these black holes by using approximate analytical and accurate numerical metric functions. We share the \emph{Mathematica®} code \cite{package} which allows one to find an appropriate analytical form of the metric for any couplings and values of parameters. Scalarization increases the radius of the black-hole shadow for all the considered coupling functions.
\end{abstract}
\maketitle

\section{Introduction}

Black holes in General Relativity possess a remarkable property: they can be fully described by only a few parameters, such as mass, angular momentum, and electric charge.
Absence of other charges, for example, scalar ones, is guaranteed by the so-called no-hair theorem \cite{Carter:1971zc,Robinson:1975bv} (see e.~g.~\cite{Chrusciel:2012jk} for review). A qualitatively different situation occurs when the scalar field is nonminimally coupled either to the gravitational sector with higher curvature corrections \cite{Doneva:2017bvd,Silva:2017uqg,Antoniou:2017hxj,Minamitsuji:2018xde,Cunha:2019dwb} or to the electromagnetic field \cite{Doneva:2010ke,Herdeiro:2018wub,Fernandes:2019rez,Astefanesei:2019pfq}. In these cases in some range of parameters of the system, the black hole acquires a scalar hair \cite{Myung:2018vug}, which was called \emph{spontaneous scalarization}. At the same time the above scalarized black-hole solutions are asymptotically flat and represent alternative models for black holes. Current experiments in the electromagnetic \cite{Akiyama:2019cqa,Akiyama:2019bqs,Goddi:2017pfy} and gravitational \cite{Abbott:2016blz,LIGOScientific:2018mvr} spectra do not allow one to determine the black-hole geometry with sufficient accuracy in order to single out the Einstein theory of gravity. Therefore, a broad parametric freedom remains for alternative theories \cite{Konoplya:2016pmh,Berti:2018vdi}, making the scalarized black holes interesting candidates for testing the no-hair theorem.

Four dimensional black-hole metrics with scalarization have been obtained only numerically, what seriously constrains the variety of tools which can be applied to study these solutions. In absence of an exact analytical solution, the analytical approximation for a metric with the controlled accuracy can remedy the situation.  A general approach based on a convergent procedure for finding such an analytical approximation was suggested in \cite{Rezzolla:2014mua} for spherical symmetry and further extended in \cite{Konoplya:2016jvv} for arbitrary axially symmetric black holes. The approach for spherical spacetimes is based on the continued fraction expansion of the metric near the event horizon in terms of a compact coordinate and matching this expansion with the post-Newtonian expansion in the far region. It was shown that the parametrization, for both spherical and axial spacetimes, usually converges quickly \cite{Younsi:2016azx}. As a rule, a few orders of expansion are sufficient for getting reasonable approximations for the metrics, so that one can compute various physical effects (quasinormal modes, particle motion, Hawking radiation, accretion etc. \cite{Kokkotas:2017zwt,Kokkotas:2017ymc,Nampalliwar:2018iru,Zinhailo:2018ska,Konoplya:2019ppy,Konoplya:2019hml}) in the black-hole background with negligible error due to the replacement of an accurate numerical black-hole solution by an approximate analytical one.

Here we will construct an analytical approximation for spherically symmetric black holes in the Einstein-Maxwell theory endowed with a scalar field which is nonminimally coupled
to the electromagnetic one. Various couplings of the scalar field are considered here and once the coupling and the physical parameters of the black hole are chosen, the analytical approximation for the metric and scalar and electromagnetic fields can be constructed. In order to understand how well the analytical metric at a given order of the expansion approximates the accurate numerical solution we calculate radii of shadows for numerical and analytical black-hole metrics.

The paper is organized as follows. In Sec.~\ref{sec:model} we give the basic information on the Einstein-Maxwell-scalar black holes at various couplings. Sec.~\ref{sec:numeric} describes the numerical solution and range of parameters for the black holes. Sec.~\ref{sec:parametrization} is devoted to application of the general parametrization \cite{Rezzolla:2014mua} to the case of the Einstein-Maxwell-scalar black holes. In Sec.~\ref{sec:shadows} we calculate the radii of shadows for the parametrized black holes and, via comparison with those for accurate numerical solutions, make conclusions on the accuracy of our analytical form at various orders of the continued fraction expansion. Finally, in Sec.~\ref{sec:conclusions} we summarize the obtained results and discuss some open questions.

\section{Einstein-Maxwell-scalar model}\label{sec:model}

We consider spherically symmetric black holes which appear in a family of Einstein-Maxwell-scalar models, described by the action
\begin{equation}\label{action}
\mathcal{S}=\int d^4 x \sqrt{-g}\left(R-2g^{\mu\nu}\partial_\mu \phi \partial_\nu \phi - f(\phi)F_{\mu\nu}F^{\mu\nu}\right),
\end{equation}
where $F_{\mu\nu}=\partial_\mu A_\nu-\partial_\nu A_\mu$ is the Maxwell tensor and $g$ is the determinant of the metric tensor $g_{\mu\nu}$.

The electric field is described by the potential
\begin{equation}\label{Maxwellfield}
A_\mu dx^\mu=V(r)dr,
\end{equation}
and the metric tensor is given by the line element
\begin{equation}\label{metric}
 ds^2 = - N(r)e^{-2\delta (r)}dt^2+\frac{dr^2}{N(r)}+r^2 d\sigma^2.
\end{equation}
The line element of a unit sphere is defines as
$$d\sigma^2 =d\theta ^2 +\sin^2\theta d \varphi ^2.$$

By substituting this ansatz for the metric tensor and electric field into the field equations one can see that
the functions $N(r)$, $\delta(r)$, $\phi(r)$, and $V(r)$ satisfy \cite{Fernandes:2019rez}
\begin{subequations}\label{eqs}
\begin{eqnarray}	
   N' &=& \frac{1}{r}\left(1-N\right)-\frac{Q^2}{r^3f(\phi)}-r(\phi')^2N,\label{Neq}\\
   \Big(r^2 N \phi ' \Big) ' &=& -\frac{f'(\phi)Q^2}{2f^2(\phi)r^2}-r^3(\phi')^3N,\label{phieq}\\
   \delta '&=& -r(\phi')^2,\label{deltaeq}\\
V' &=& \frac{Q}{f(\phi) r^2} e^{-\delta},\label{Veq}
\end{eqnarray}
\end{subequations}
where $Q$ is a constant of integration corresponding to the electric charge.

\begin{subequations}\label{ascond}
We assume that the asymptotic is Minkowskian and the time is measured by the coordinate $t$, so that,
\begin{equation}
\lim_{r\to\infty}N=1, \qquad \lim_{r\to\infty}\delta=0.
\end{equation}
We also suppose that
\begin{equation}
\lim_{r\to\infty}\phi=0,\qquad \lim_{r\to\infty}V=0.
\end{equation}
\end{subequations}

If $f'(0)=0$,  equations (\ref{eqs}) have the nonscalarized Reissner-Nordström solution for $\delta=\phi=0$,
\begin{equation}
N=1-\frac{2M}{r}+\frac{Q^2}{r^2f(0)}, \qquad V=\frac{Q}{r^2f(0)}.
\end{equation}
Thus, without loss of generality, in order to identify the arbitrary constant $Q$ with the electric charge, we assume that $f(0)=1$.

Further we shall consider only those solutions of Eq.~(\ref{eqs}), for which the scalar field is positive-definite ($\phi>0$) everywhere.\footnote{Note, that the solutions, for which $\phi$ changes its sign, were proven to be unstable \cite{Myung:2018jvi}.}
Hence, for simplicity, we also assume that the ad~hoc function $f(\phi)$ is monotonously growing for all allowed positive values of $\phi<\phi_{max}$. This assumption automatically satisfies the constrains coming from the Bekenstein-type identities which were discussed in \cite{Fernandes:2019rez}.
The above conditions are satisfied for all the coupling types considered in \cite{Fernandes:2019rez}, which are
\begin{enumerate}
 \item[1)] an exponential coupling, $f(\phi) = e^{-\alpha\phi^2}$, first considered in this context in~\cite{Herdeiro:2018wub};
 \item[2)] a hyperbolic cosine coupling, $$f(\phi) = \cosh({\sqrt{-2\alpha}\phi});$$
 \item[3)] a power coupling, $f(\phi) = 1-\alpha \phi^2$;
 \item[4)] a fractional coupling, $f(\phi) = \frac{1}{1+\alpha\phi^2}$.
\end{enumerate}
Here $\alpha<0$ is a dimensionless constant, so that
$$f(\phi)=1-\alpha\phi^2+{\cal O}(\phi^4).$$

In addition, we can consider other coupling types, satisfying the same condition:
\begin{enumerate}
 \item[5)] $f(\phi) = e^{\dfrac{e^{-\alpha\beta^2\phi^2}-1}{\beta^2}}$;
 \item[6)] $f(\phi) = \left(1-\alpha \frac{\phi^2}{n}\right)^n$;
 \item[7)] $f(\phi) = \cosh({\sqrt{-\frac{2\alpha}{n}}\phi})^n$; etc.
\end{enumerate}

Now we are in a position to consider the numerical solutions representing scalarized black holes for the above couplings of a scalar field.

\section{Numerical solution for a scalarized black hole}\label{sec:numeric}

Following \cite{Herdeiro:2018wub}, we search for the numerical black-hole solution of the equations (\ref{eqs}) allowing for the scalar hair, i.~e. we assume that at the event horizon is located at $r_0$, so that $N(r_0)=0$, and consider $\phi(r_0)=P_0>0$. From (\ref{phieq}) and (\ref{Neq}) we find that
\begin{equation}
\phi'(r_0)=-\frac{f'(P_0)}{2f(P_0)}\frac{Q^2}{f(P_0)r_0^2-Q^2}\leq0,
\end{equation}
being nonpositive, since $f(\phi)>0$ is monotonously growing and the Hawking temperature at the horizon
$$T_H=\frac{N'(r_0)}{4\pi}e^{-\delta(r_0)}=\frac{1}{4\pi r_0}\left(1-\frac{Q^2}{f(P_0)r_0^2}\right)e^{-\delta_0}>0.$$

With these initial conditions at the horizon we search for the solutions of (\ref{Neq}) and (\ref{phieq}), for which $\phi>0$ for any $r_0\leq r<\infty$ and $\phi=0$ at spatial infinity. Using the shooting method \cite{shooting}, we determine the corresponding charge $Q$ in the interval\footnote{In principle, it is possible to solve the inverse problem: Using the shooting method, one could find the value of $P_0$ for any given $Q$. However, such an approach would require inverting the inequality (\ref{Qlim}), which leads to unnecessary complications.}
\begin{equation}\label{Qlim}
  0<Q<r_0\sqrt{f(P_0)}.
\end{equation}

Once the parameters $P_0$ and $Q$ are fixed, we can solve the remaining equations (\ref{deltaeq})~and~(\ref{Veq}) numerically. Since the solutions of (\ref{deltaeq}) and (\ref{Veq}) for $\delta(r)$ and $V(r)$ differ by an arbitrary constant, we do not need to shoot for the parameters $\delta_0=\delta(r_0)$ and $V_0=V(r_0)$ in order to satisfy the asymptotic conditions (\ref{ascond}). For simplicity of the algorithm, we find numerically the shifted functions,
\begin{equation}\label{shiftedfunc}
\bar{\delta}(r)\equiv\delta(r)-\delta_0, \qquad \bar{V}(r)\equiv e^{\delta_0}V(r)-V_0,
\end{equation}
which are equal to zero at the event horizon.

In order to solve the differential equations we rewrite (\ref{eqs}) in terms of the compact coordinate
\begin{equation}\label{compactcoord}
x=1-\frac{r_0}{r},\qquad 0\leq x<1,
\end{equation}
substitute the two equivalent first-order equations instead of (\ref{phieq}) and use the LSODA method \cite{LSODE} implemented in \emph{Wolfram ®Mathematica} for solving the equations in the interval $x_0\leq x\leq x_1$. The initial conditions are imposed at $x=x_0>0$ using a Maclaurin series expansion up to the sixth order for the functions $N(x)$, $\phi(x)$, $\delta(x)$, and $V(x)$, and the point $x_0\ll1$ is chosen in order to match the desired numerical precision. The final point $x_1\lessapprox1$ is fixed by the standard stiffness detection of the \emph{Mathematica® NDSolve} function.

In order to determine the asymptotic parameters, we compare the numerical solution with the asymptotic expansion
\begin{subequations}\label{asexp}
\begin{eqnarray}
  N(r) &=& 1-\frac{2M}{r}+\frac{Q^2+S^2}{r^2}+{\cal O}\left(\frac{1}{r^3}\right), \label{asN}\\
  \delta(r) &=& \frac{S^2}{2r^2}+{\cal O}\left(\frac{1}{r^3}\right), \label{asD}\\
  \phi(r) &=& \frac{S}{r}+{\cal O}\left(\frac{1}{r^2}\right), \label{asP} \\
  V(r) &=& \frac{Q}{r}+{\cal O}\left(\frac{1}{r^2}\right). \label{asV}
\end{eqnarray}
\end{subequations}

The parameters satisfy the following relation \cite{Herdeiro:2018wub},
\begin{eqnarray}\label{videntity}
M^2 + S^2 &=& Q^2 + 4\pi^2 r_0^4~ T_H^2 \\\nonumber &=&Q^2 + \frac{r_0^2}{4} \left(1-\frac{Q^2}{f(P_0)r_0^2}\right)^2e^{-2\delta_0}.
\end{eqnarray}
We have checked that (\ref{videntity}) is satisfied with good accuracy for the numerically calculated parameters. It turns out though, that, for given numerical precision, this relation provides a better accuracy for the asymptotic mass than a numerical extrapolation of $N'(x)$ at $x\to1$. That is why we use (\ref{videntity}) to calculate the black-hole mass $M$.

\section{Analytic representation}\label{sec:parametrization}

\begin{figure*}
\resizebox{\linewidth}{!}{\includegraphics*{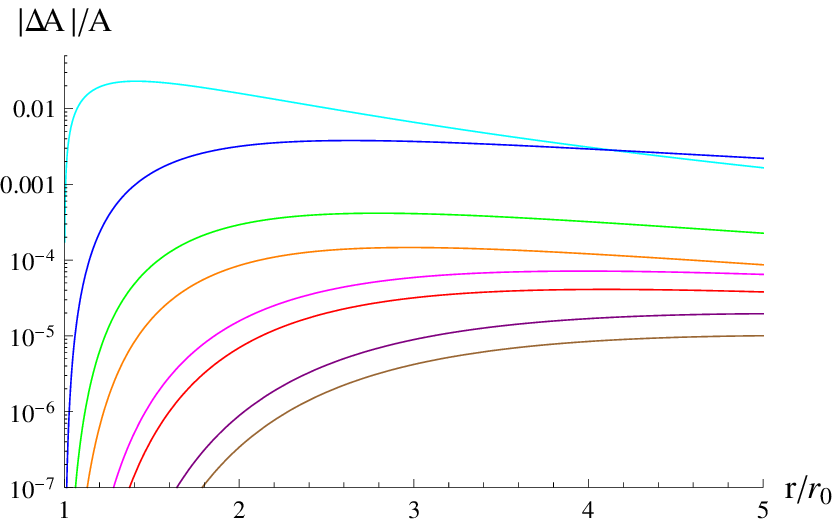}\includegraphics*{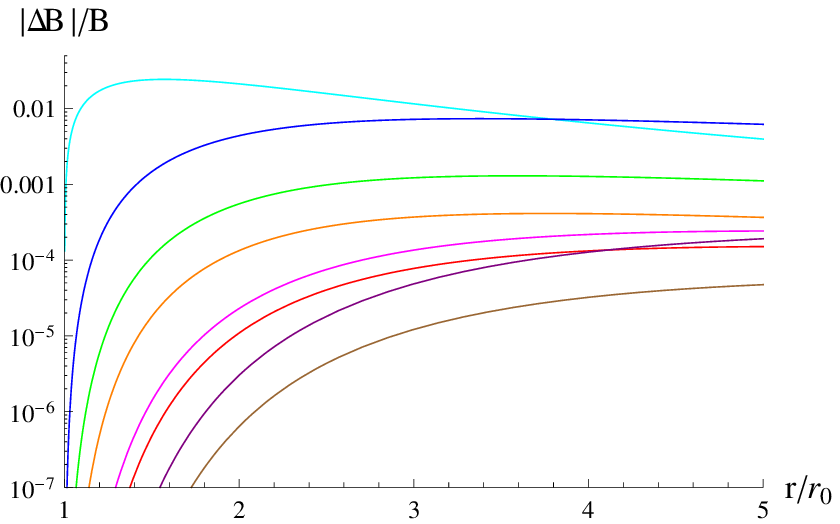}}
\resizebox{\linewidth}{!}{\includegraphics*{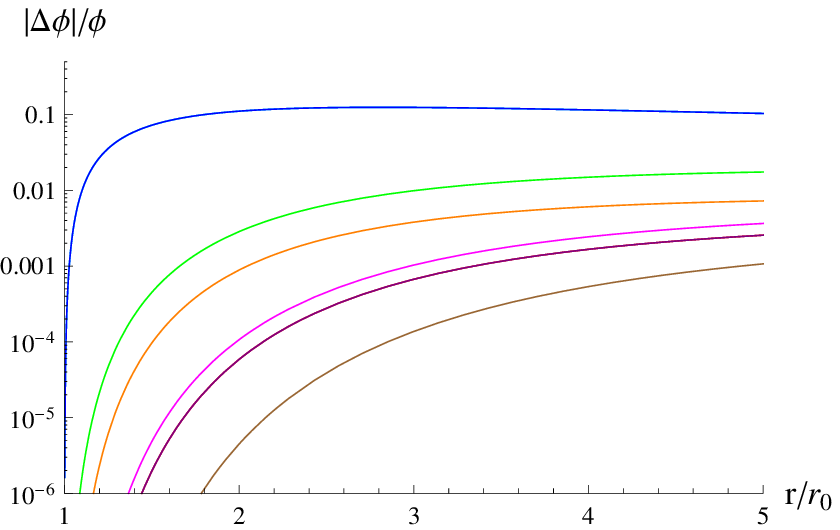}\includegraphics*{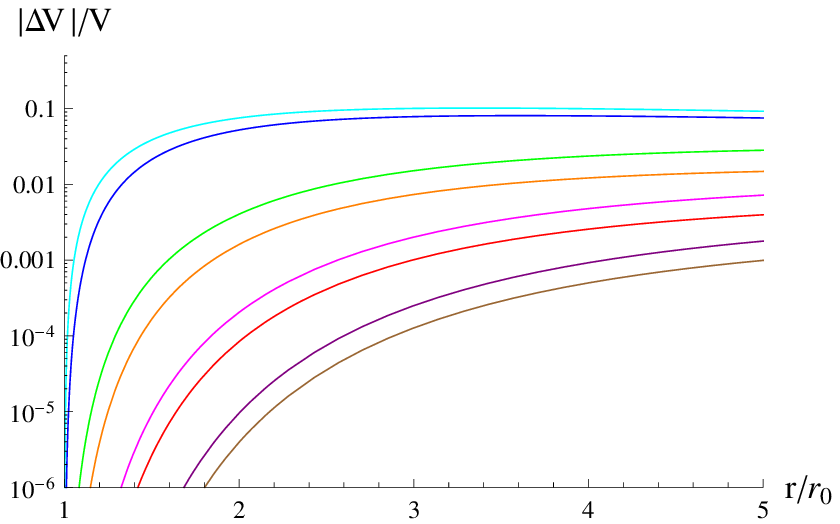}}
\caption{Relative error of the analytical approximation of order (from top to bottom): 2 (cyan), 3 (blue), 4 (green), 5 (orange), 6 (magenta), 7 (red), 8 (purple), 9 (brown) for the scalarized black hole with the coupling $f(\phi)=e^{10\phi^2}$, $P_0=0.5$, $Q\approx1.104M$, $S\approx0.748M$.}\label{fig:metric}
\end{figure*}

Following \cite{Rezzolla:2014mua}, we represent the black-hole metric as,
\begin{eqnarray}\label{parammetric}
  ds^2&=&-\left(1-\frac{r_0}{r}\right)A\left(1-\frac{r_0}{r}\right)dt^2\\\nonumber&&+\dfrac{B^2\left(1-\frac{r_0}{r}\right)dr^2}{\left(1-\frac{r_0}{r}\right)A\left(1-\frac{r_0}{r}\right)}+r^2d\sigma^2,\\\nonumber
\end{eqnarray}
where the functions $A(x)$ and $B(x)$
$$A(x)\simeq\frac{N(x)}{x}e^{-2\delta(x)},\qquad B(x)\simeq e^{-\delta(x)},$$
are finite everywhere for $0\leq x\leq1$ and represented as
\begin{eqnarray}
A(x)&=&1-\epsilon (1-x)+(a_0-\epsilon)(1-x)^2+{\tilde A}(x)(1-x)^3,\nonumber\\
\label{asympfix}
B(x)&=&1+b_0(1-x)+{\tilde B}(x)(1-x)^2,
\end{eqnarray}
%%%%%%
where ${\tilde A}(x)$ and ${\tilde B}(x)$ are given in terms of the continued fractions, in order to describe the metric near the event horizon $x=0$:
%%%%%
\begin{align}\nonumber
{\tilde A}(x)=\frac{a_1}{\displaystyle 1+\frac{\displaystyle
    a_2x}{\displaystyle 1+\frac{\displaystyle a_3x}{\displaystyle
      1+\frac{\displaystyle a_4x}{\displaystyle
      1+\ldots}}}}\,,\\\label{contfrac}
%%%%%%
{\tilde B}(x)=\frac{b_1}{\displaystyle 1+\frac{\displaystyle
    b_2x}{\displaystyle 1+\frac{\displaystyle b_3x}{\displaystyle
      1+\frac{\displaystyle b_4x}{\displaystyle
      1+\ldots}}}}\,.
\end{align}

The coefficients $a_0$, $b_0$, and $\epsilon$ are fixed by comparing asymptotic expansions (\ref{asexp}) and (\ref{asympfix}) as follows
\begin{equation}
\epsilon=\frac{2M-r_0}{r_0},\qquad a_0=\frac{Q^2}{r_0^2},\qquad b_0=0.
\end{equation}

In a similar manner we introduce the analytic representation for the fields $\phi(x)$ and $V(x)$,
\begin{eqnarray}
\phi(x)&\simeq& p_0(1-x)+{\tilde P}(x)(1-x)^2, \label{asympfixP}\\
V(x)&\simeq& v_0(1-x)+{\tilde V}(x)(1-x)^2, \label{asympfixV}
\end{eqnarray}
where $p_0$ and $v_0$ are fixed by comparison with (\ref{asP}) and (\ref{asV}), respectively, as
\begin{equation}
p_0=\frac{S}{r_0},\qquad v_0=\frac{Q}{r_0},
\end{equation}
and
\begin{align}\nonumber
{\tilde P}(x)=\frac{p_1}{\displaystyle 1+\frac{\displaystyle
    p_2x}{\displaystyle 1+\frac{\displaystyle p_3x}{\displaystyle
      1+\frac{\displaystyle p_4x}{\displaystyle
      1+\ldots}}}}\,,\\\label{contfracfields}
%%%%%%
{\tilde V}(x)=\frac{v_1}{\displaystyle 1+\frac{\displaystyle
    v_2x}{\displaystyle 1+\frac{\displaystyle v_3x}{\displaystyle
      1+\frac{\displaystyle v_4x}{\displaystyle
      1+\ldots}}}}\,.
\end{align}

Expanding (\ref{asympfix}), (\ref{asympfixP}), and (\ref{asympfixV}) near the event horizon ($x=0$) and substituting into (\ref{eqs}) we calculate numerically the other coefficients $a_1, a_2, a_3,\ldots$, $b_1,b_2,b_3,\ldots$, $p_1,p_2,p_3,\ldots$, $v_1,v_2,v_3,\ldots$ up to any given order. In particular,
\begin{eqnarray}
1+b_1=B(0)=e^{-\delta_0},\quad p_0+p_1=\phi(0)=P_0.\qquad
\end{eqnarray}

Since we use (\ref{videntity}) to calculate the black hole mass, the following relation holds
\begin{eqnarray}
(\epsilon+1)^2&=&\frac{4Q^2-4S^2}{r_0^2} + \left(1-\frac{Q^2}{f(P_0)r_0^2}\right)^2e^{-2\delta_0}\\\nonumber
&=&4v_0^2-4p_0^2 + \left(1-\frac{a_0}{f(p_0+p_1)}\right)^2(1+b_1)^2.
\end{eqnarray}

The hierarchy of the near-horizon coefficients introduced through the continued fractions (\ref{contfrac}) and (\ref{contfracfields}) implies that for finite floating point size of mantissa we are able to calculate finite number of the meaningful coefficients. However, by increasing the numerical precision one can find as many coefficients as needed. We have used the \emph{®Mathematica's} powerful arbitrary precision arithmetics and built-in precision control for the calculation of the coefficients.

On Fig.~\ref{fig:metric} we show convergence of the above procedure. By increasing order of the continued fraction we are able to approximate all the functions as good as necessary. One should note that the approximations of certain orders are not always possible to obtain in a consistent manner by setting the higher-order term equal to zero. The reason is that the truncated continued fraction can have singular points outside the horizon (see discussion in Sec.~IV of \cite{Konoplya:2016jvv}). For simplicity, we do not consider approximations of such problematic orders and generate the approximated functions of the lower orders instead. That is why on Fig.~\ref{fig:metric} we see that the approximated function for the scalar field at the third order coincides with the one at the second order. The higher-order approximations amend the problem and the sequence of approximations converges.

The \emph{Mathematica® package} with a numerical code for the calculation of the coefficients and analytic representations for all the functions, (\ref{asympfix}), (\ref{asympfixP}) and (\ref{asympfixV}), and a sample \emph{notebook} ``SBHDemo.nb'' can be found in \cite{package}. In the appendix, as examples, we write down the values of the coefficients of the parametrization for a few fixed couplings and values of physical parameters of the black hole.

\section{Shadows}\label{sec:shadows}
In order to find out how the found approximation is good, one needs to compare some observable, gauge invariant quantity obtained for the approximate and numerical metrics. Such a simple and meaningful quantity is the radius of the black-hole shadow \cite{Bambi:2008jg,Bambi:2010hf,Amarilla:2010zq,Johannsen:2010ru}, which has recently been studied in a number of works (see e.~g.~\cite{Konoplya:2019fpy,Wang:2019tjc,Bisnovatyi-Kogan:2018vxl,Amir:2017slq,Xu:2018mkl,Wang:2017hjl,Tsukamoto:2017fxq,Abdujabbarov:2017pfw,Wang:2018eui,Hennigar:2018hza,Shaikh:2018kfv,Zhu:2019ura,Konoplya:2019sns,Contreras:2019cmf,Perlick:2015vta,Ovgun:2019jdo,Long:2019nox,Vagnozzi:2019apd,Contreras:2019nih,Held:2019xde,Dokuchaev:2018fze,Jusufi:2019nrn} and references therein).
For the spherically symmetric black hole (\ref{parammetric}) the shadow radius, visible by a remote observer, is
\begin{equation}\label{shadow}
R=\frac{r_p}{\sqrt{F(r_p)}},
\end{equation}
where
$$F(r)=\left(1-\frac{r_0}{r}\right)A\left(1-\frac{r_0}{r}\right),$$
and $r_p$ is the coordinate of the circular photon orbit, satisfying,
\begin{eqnarray}
  ds^2 &=& -F(r_p)dt^2+r_p^2d\sigma^2 = 0, \\\nonumber
  d^2r &=& \left(-\dfrac{F'(r_p)}{2}dt^2+r_pd\sigma^2\right)\dfrac{F(r_p)}{B^2\left(1-r_0/r_p\right)} = 0,
\end{eqnarray}
or, equivalently,
$$r_pF'(r_p)=2F(r_p).$$
Hence, the shadow radius is given by the minimal value of the function $h(r)$,
\begin{equation}\label{minfunc}
  h^2(r)=\frac{r^2}{F(r)}=\frac{r^3}{(r-r_0)A(1-r_0/r)}.
\end{equation}

The minimum of $h(r)$ can be easily calculated numerically once we have an analytical approximation for the function $A(x)$ in the form of a rational functions. Since $h^2(r)$ is a rational function of $r$, the corresponding minimum is the smallest real, larger than $r_0$, root of some polynomial.

\begin{figure}
\resizebox{\linewidth}{!}{\includegraphics*{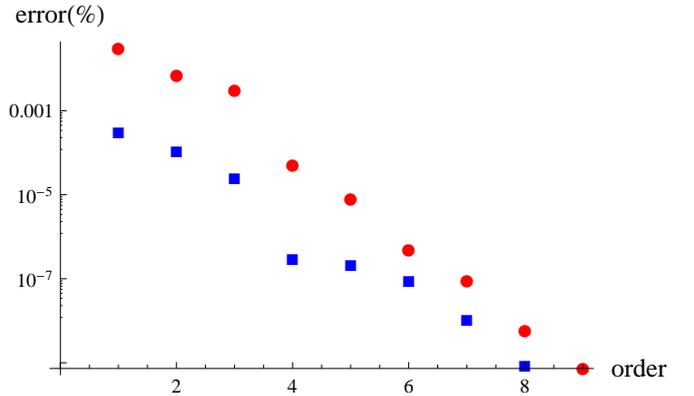}}
\caption{Semi-logarithmic plot for the relative error for the shadow radius as a function of the approximation order for the scalarized black hole with the coupling $f(\phi)=1+1000\phi^2$: $P_0=0.026$, $Q\approx0.105M$, $S\approx0.0417M$ (blue squares) and $P_0=0.094$, $Q\approx0.44M$, $S\approx0.365M$ (red dots).}\label{fig:convergence}
\end{figure}

First, we compare the result of numerical minimizing of the function (\ref{minfunc}) obtained from the numerical integration with the one obtained using analytical approximations of various orders. On Fig.~\ref{fig:convergence} we see that convergence is fast for any charge of the scalarized black hole. We notice that, for the particular problem, one can find the accurate value of the shadow size using numerical solution only by minimizing the function (\ref{minfunc}). That is why the shadow size is a good test for convergence of our method. However, the numerical solution is not useful for the tasks, which require higher derivatives, such as, for example, calculations of quasinormal modes \cite{Konoplya:2011qq} in the frequency domain. For instance, the WKB quasinormal frequencies reported in \cite{Cai:2015fia} proved to be inaccurate because of the enormous numerical error accumulated when taking higher derivatives from the numerically given metric function. On the contrary, the analytical approximation used in \cite{Zinhailo:2018ska} led to reliable results for the quasinormal modes. Therefore, our procedure for construction of convergent analytical approximations for the metric functions and fields opens a window for such analyses.

\begin{figure}
\resizebox{\linewidth}{!}{\includegraphics*{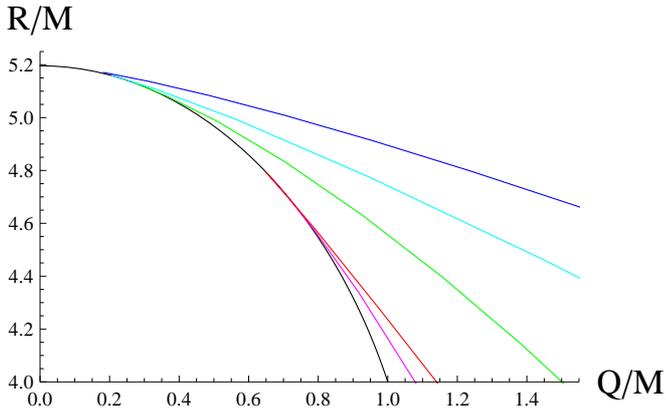}}
\caption{Shadow size for the charged Reissner-Nordström black hole (black, bottom) and scalarized black holes with various couplings (from bottom to top): $f(\phi)=e^{10\phi^2}$ (magenta), $f(\phi)=e^{e^{10\phi^2}-1}$ (red), $f(\phi)=1+100\phi^2$ (green), $f(\phi)=(1+100\phi^2)^2$ (cyan), and $f(\phi)=1+1000\phi^2$ (blue, top).}\label{fig:shadows}
\end{figure}

Finally, on Fig.~\ref{fig:shadows} we show the shadow of the Reissner-Nordström black hole with the ones of scalarized black holes. We see that scalarization increases visible size of the black hole for all the considered coupling functions. The faster coupling function grows the larger deviation from the Reissner-Nordström black hole is and the smaller charge allows for the scalarized branch to appear.

The \emph{Mathematica® notebook} with a numerical code for the calculation of shadows ``SBHShadows.nb'' can be found in \cite{package}.

\section{Conclusions}\label{sec:conclusions}

Here we have found analytical approximations for the black hole metric in the Einstein-Maxwell-scalar theory, once the coupling of the scalar field and physical parameters of the system are chosen. In the general case, when none of the parameters are fixed, the fitting of the parametrization to the numerical solution is a time consuming problem which, in principle, could be solved in the future. The obtained analytical approximations are applied here for calculation of shadows cast by scalarized black holes. It has been found that the scalarization increases the radius of the shadow for every coupling under consideration. The continued fraction expansion, which we used for finding the analytical form of the metric, converges quickly, showing reasonable accuracy already at the second order. We share with readers the \emph{Mathematica®} codes which makes it possible both to find analytical forms of the approximate metric functions for any desired values of the parameters as well as to calculate radii of shadows for each case. The analytical approximations generated by our method are ready to use for further study of the Einstein-Maxwell-scalar black holes and phenomena in their vicinity, such as particle motion, quasinormal ringing, stability etc.

\begin{acknowledgments}
A.~Z. was supported by Conselho Nacional de Desenvolvimento Científico e Tecnológico (CNPq) and the grant 19-03950S of Czech Science Foundation (GAČR). R.~K. was supported by the grant 19-03950S of Czech Science Foundation (GAČR). This publication has been prepared with the support of the ``RUDN University Program 5-100''.
\end{acknowledgments}

\appendix
\section*{Appendix: Coefficients of the analytic representation}

\begin{table*}
\begin{tabular}{|l|r|r|r|r|r|r|r|r|r|r|}
\hline
 $P_0     $ & \multicolumn {1}{c|}{$0.01$} & \multicolumn {1}{c|}{$0.02$} & \multicolumn {1}{c|}{$0.03$} & \multicolumn {1}{c|}{$0.04$} & \multicolumn {1}{c|}{$0.05$} & \multicolumn {1}{c|}{$0.06$} & \multicolumn {1}{c|}{$0.07$} & \multicolumn {1}{c|}{$0.08$} & \multicolumn {1}{c|}{$0.09$} & \multicolumn {1}{c|}{$0.10$} \\
\hline
 $r_0/M   $ & $ 1.996783$ & $ 1.995736$ & $ 1.993844$ & $   1.990967$ & $ 1.986978$ & $ 1.981777$ & $ 1.975285$ & $ 1.967446$ & $ 1.958222$ & $ 1.947591$ \\
 $Q/M     $ & $ 0.080199$ & $ 0.093082$ & $ 0.114516$ & $   0.144319$ & $ 0.182262$ & $ 0.228122$ & $ 0.281678$ & $ 0.342699$ & $ 0.410931$ & $ 0.486093$ \\
 $S/M     $ & $ 0.013038$ & $ 0.029333$ & $ 0.051210$ & $   0.079943$ & $ 0.116111$ & $ 0.159910$ & $ 0.211336$ & $ 0.270269$ & $ 0.336509$ & $ 0.409803$ \\
 $\epsilon$ & $ 0.001611$ & $ 0.002137$ & $ 0.003088$ & $   0.004537$ & $ 0.006554$ & $ 0.009195$ & $ 0.012512$ & $ 0.016546$ & $ 0.021335$ & $ 0.026909$ \\
 $a_0     $ & $ 0.001613$ & $ 0.002175$ & $ 0.003299$ & $   0.005254$ & $ 0.008414$ & $ 0.013250$ & $ 0.020335$ & $ 0.030340$ & $ 0.044037$ & $ 0.062296$ \\
 $a_1     $ & $ 0.000015$ & $ 0.000089$ & $ 0.000236$ & $   0.000361$ & $ 0.000240$ & $-0.000477$ & $-0.002260$ & $-0.005695$ & $-0.011477$ & $-0.020411$ \\
 $a_2     $ & $ 0.381378$ & $-0.097074$ & $-0.651319$ & $  -1.332591$ & $-3.652014$ & $ 2.221766$ & $ 0.275885$ & $-0.171688$ & $-0.383614$ & $-0.510693$ \\
 $a_3     $ & $-0.532212$ & $ 2.473512$ & $ 0.509826$ & $   0.684250$ & $ 2.635464$ & $-3.715004$ & $-3.562145$ & $ 2.006467$ & $ 0.428381$ & $ 0.176736$ \\
 $a_4     $ & $ 0.386404$ & $-2.748436$ & $-0.610322$ & $  -0.282888$ & $-0.106041$ & $ 0.199664$ & $ 1.856321$ & $-3.373555$ & $-1.674276$ & $-1.376034$ \\
 $b_1     $ & $-0.000064$ & $-0.000228$ & $-0.000453$ & $  -0.000721$ & $-0.001028$ & $-0.001374$ & $-0.001759$ & $-0.002185$ & $-0.002652$ & $-0.003160$ \\
 $b_2     $ & $ 0.802289$ & $ 0.169424$ & $-0.335728$ & $  -0.654802$ & $ -0.84816$ & $-0.967782$ & $-1.044447$ & $-1.095293$ & $-1.130015$ & $-1.154292$ \\
 $b_3     $ & $-0.814279$ & $-2.675663$ & $ 0.862733$ & $   0.322884$ & $ 0.215554$ & $ 0.180470$ & $ 0.166249$ & $ 0.159450$ & $ 0.155626$ & $ 0.153104$ \\
 $b_4     $ & $ 0.306433$ & $ 2.049811$ & $-1.507134$ & $  -0.840876$ & $-0.587038$ & $-0.454977$ & $-0.388166$ & $-0.354325$ & $-0.336566$ & $-0.326712$ \\
 $p_0     $ & $ 0.006530$ & $ 0.014698$ & $ 0.025684$ & $   0.040153$ & $ 0.058436$ & $ 0.080690$ & $ 0.106990$ & $ 0.137370$ & $ 0.171844$ & $ 0.210415$ \\
 $p_1     $ & $ 0.003470$ & $ 0.005302$ & $ 0.004316$ & $  -0.000153$ & $-0.008436$ & $-0.020690$ & $-0.036990$ & $-0.057370$ & $-0.081844$ & $-0.110415$ \\
 $p_2     $ & $-0.034343$ & $-0.579136$ & $-1.588200$ & $  56.836876$ & $ 0.846199$ & $ 0.078779$ & $-0.216913$ & $-0.381338$ & $-0.487965$ & $-0.563245$ \\
 $p_3     $ & $ 3.286144$ & $ 0.589416$ & $ 1.036123$ & $ -57.836519$ & $-2.489250$ & $-7.057205$ & $ 1.091320$ & $ 0.309945$ & $ 0.126922$ & $ 0.055830$ \\
 $p_4     $ & $-3.176566$ & $-0.498698$ & $-0.210475$ & $   0.005762$ & $ 0.440551$ & $ 5.606377$ & $-2.392389$ & $-1.591872$ & $-1.479008$ & $-1.610655$ \\
 $v_0     $ & $ 0.040164$ & $ 0.046640$ & $ 0.057435$ & $   0.072487$ & $ 0.091728$ & $ 0.115110$ & $ 0.142601$ & $ 0.174185$ & $ 0.209849$ & $ 0.249587$ \\
 $v_1     $ & $-0.001063$ & $-0.004639$ & $-0.011506$ & $  -0.022287$ & $-0.037312$ & $-0.056707$ & $-0.080499$ & $-0.108673$ & $-0.141198$ & $-0.178040$ \\
 $v_2     $ & $ 1.436696$ & $ 0.874088$ & $ 0.365699$ & $   0.002384$ & $ -0.24326$ & $-0.410771$ & $-0.528260$ & $-0.613295$ & $-0.676682$ & $-0.725179$ \\
 $v_3     $ & $-1.229244$ & $-1.407492$ & $-1.983863$ & $-169.935381$ & $ 0.933804$ & $ 0.317194$ & $ 0.144572$ & $ 0.073922$ & $ 0.039709$ & $ 0.021458$ \\
 $v_4     $ & $ 0.181564$ & $ 0.301386$ & $ 0.835521$ & $ 168.778908$ & $-2.088528$ & $-1.477205$ & $-1.325626$ & $-1.299029$ & $-1.343519$ & $-1.463914$ \\
\hline
\end{tabular}
\caption{List of coefficients of the parametrization for the polynomial coupling $f(\phi)=1+1000\phi^2$.}\label{tabl:poly}
\end{table*}

\begin{table*}
\begin{tabular}{|l|r|r|r|r|r|r|r|r|r|r|}
\hline
$f(\phi)$ & \multicolumn {2}{c|}{$e^{2\phi^2}$} & \multicolumn {3}{c|}{$e^{10\phi^2}$} & \multicolumn {3}{c|}{$1+100\phi^2$} & \multicolumn {2}{c|}{$1/(1-20\phi^2$)} \\
\hline
 $P_0     $ & \multicolumn {1}{c|}{$0.1$} & \multicolumn {1}{c|}{$0.5$} & \multicolumn {1}{c|}{$0.1$} & \multicolumn {1}{c|}{$0.3$} & \multicolumn {1}{c|}{$0.5$} & \multicolumn {1}{c|}{$0.1$} & \multicolumn {1}{c|}{$0.3$} & \multicolumn {1}{c|}{$0.5$} & \multicolumn {1}{c|}{$0.1$} & \multicolumn {1}{c|}{$0.2$} \\
\hline
 $r_0/M   $ & $  1.299434$ & $  1.170318$ & $ 1.748684$ & $ 1.640257$ & $ 1.365362$ & $ 1.936884$ & $ 1.621922$ & $ 1.163725$ & $ 1.868478$ & $ 1.866893$ \\
 $Q/M     $ & $  0.954134$ & $  0.995127$ & $ 0.663182$ & $ 0.789308$ & $ 1.103900$ & $ 0.363384$ & $ 1.154977$ & $ 1.991225$ & $ 0.495761$ & $ 0.499028$ \\
 $S/M     $ & $  0.049610$ & $  0.251754$ & $ 0.101502$ & $ 0.348868$ & $ 0.747986$ & $ 0.168097$ & $ 0.944931$ & $ 1.787795$ & $ 0.111650$ & $ 0.226239$ \\
 $\epsilon$ & $  0.539132$ & $  0.708938$ & $ 0.143717$ & $ 0.219321$ & $ 0.464813$ & $ 0.032586$ & $ 0.233105$ & $ 0.718619$ & $ 0.070390$ & $ 0.071299$ \\
 $a_0     $ & $  0.539151$ & $  0.723018$ & $ 0.143828$ & $ 0.231563$ & $ 0.653675$ & $ 0.035199$ & $ 0.507092$ & $ 2.927789$ & $ 0.070399$ & $ 0.071451$ \\
 $a_1     $ & $  0.001771$ & $  0.075100$ & $ 0.001510$ & $ 0.025580$ & $ 0.048870$ & $ 0.002662$ & $-0.145615$ & $-1.720891$ & $ 0.001225$ & $ 0.005111$ \\
 $a_2     $ & $  2.044634$ & $  0.566020$ & $ 0.669896$ & $-0.264113$ & $-2.750254$ & $-0.732969$ & $-0.430299$ & $-0.780042$ & $ 0.639028$ & $ 0.559543$ \\
 $a_3     $ & $ -0.652142$ & $ -1.963133$ & $-0.448717$ & $ 1.525364$ & $ 1.804389$ & $ 0.503882$ & $ 0.315655$ & $ 0.009144$ & $-0.343249$ & $-0.461533$ \\
 $a_4     $ & $  0.549928$ & $  1.550105$ & $ 0.325538$ & $-1.738745$ & $-0.187936$ & $-0.553786$ & $-1.543958$ & $-1.506366$ & $ 0.198781$ & $ 0.374022$ \\
 $b_1     $ & $ -0.009448$ & $ -0.177060$ & $-0.006896$ & $-0.049438$ & $-0.096261$ & $-0.004956$ & $-0.028872$ & $-0.068696$ & $-0.006824$ & $-0.026601$ \\
 $b_2     $ & $  3.268155$ & $  0.835333$ & $ 1.223458$ & $-0.130796$ & $-1.245455$ & $-0.389067$ & $-1.136188$ & $-1.192663$ & $ 1.239694$ & $ 1.076335$ \\
 $b_3     $ & $ -1.348604$ & $ -2.790184$ & $-0.841009$ & $ 5.170616$ & $ 0.471154$ & $ 0.719105$ & $ 0.152821$ & $ 0.137388$ & $-0.721450$ & $-0.823531$ \\
 $b_4     $ & $  0.685638$ & $  1.854161$ & $ 0.323971$ & $-5.695685$ & $-0.629005$ & $-1.358530$ & $-0.331332$ & $-0.299246$ & $ 0.234568$ & $ 0.349675$ \\
 $p_0     $ & $  0.038178$ & $  0.215116$ & $ 0.058045$ & $ 0.212691$ & $ 0.547830$ & $ 0.086787$ & $ 0.582599$ & $ 1.536269$ & $ 0.059754$ & $ 0.121185$ \\
 $p_1     $ & $  0.061822$ & $  0.284884$ & $ 0.041955$ & $ 0.087309$ & $-0.047830$ & $ 0.013213$ & $-0.282599$ & $-1.036269$ & $ 0.040246$ & $ 0.078815$ \\
 $p_2     $ & $  1.008263$ & $ -0.013465$ & $ 0.182480$ & $-0.864917$ & $ 3.526527$ & $-1.789192$ & $-0.505497$ & $-0.752991$ & $ 0.222534$ & $ 0.141264$ \\
 $p_3     $ & $ -0.140607$ & $ 54.869151$ & $-0.572085$ & $ 0.865476$ & $-4.837234$ & $ 1.177858$ & $ 0.109794$ & $-0.017049$ & $-0.190524$ & $-0.687437$ \\
 $p_4     $ & $  0.767096$ & $-54.945175$ & $ 0.822005$ & $-0.518215$ & $ 0.137434$ & $-0.188133$ & $-1.496323$ & $ 0.028595$ & $ 0.391289$ & $ 0.975307$ \\
 $v_0     $ & $  0.734269$ & $  0.850305$ & $ 0.379246$ & $ 0.481210$ & $ 0.808502$ & $ 0.187613$ & $ 0.712104$ & $ 1.711078$ & $ 0.265329$ & $ 0.267304$ \\
 $v_1     $ & $ -0.004249$ & $ -0.117107$ & $-0.010282$ & $-0.108970$ & $-0.431529$ & $-0.040889$ & $-0.493581$ & $-1.450369$ & $-0.014415$ & $-0.058160$ \\
 $v_2     $ & $  3.022157$ & $  1.636721$ & $ 1.740190$ & $ 0.709340$ & $-0.265413$ & $ 0.305531$ & $-0.697378$ & $-0.862505$ & $ 1.781910$ & $ 1.701260$ \\
 $v_3     $ & $ -1.469266$ & $ -1.842810$ & $-1.252115$ & $-1.751660$ & $ 1.269789$ & $-2.202308$ & $ 0.032153$ & $-0.002533$ & $-1.190291$ & $-1.217074$ \\
 $v_4     $ & $  0.406281$ & $  0.580785$ & $ 0.206170$ & $ 0.613641$ & $-2.457384$ & $ 1.048253$ & $-1.366314$ & $ 0.669487$ & $ 0.150501$ & $ 0.178539$ \\
\hline
\end{tabular}
\caption{List of coefficients of the parametrization for various couplings.}\label{tabl:var}
\end{table*}

In tables~\ref{tabl:poly}~and~\ref{tabl:var} we give a few examples of values of parametrization coefficients for some fixed couplings.

\end{document}